\documentclass[aps,amsfonts]{revtex4}
\usepackage{epsfig}
\usepackage{graphicx}
\usepackage{amsmath}
\usepackage{amsbsy}
\begin{document}
\newcommand{\ee}{\end{equation}}
\newcommand{\bb}{\begin{equation}}
\newcommand{\eqb}{\begin{eqnarray}}
\newcommand{\eqf}{\end{eqnarray}}
\def\sigmavec{\mbox{\boldmath$\sigma$}}
\def\x{\mathbf{x}}
\def\p{\mathbf{p}}
\def\ho{{\mbox{\tiny{HO}}}}
\def\sc{\scriptscriptstyle}

\def\sigmavec{\mbox{\boldmath$\sigma$}}
\def\nablavec{\mbox{\boldmath$\nabla$}}

\title{Magnetic-Dipole Spin Effects in Noncommutative Quantum Mechanics}
\author{H.  Falomir$^{a}$,  J. Gamboa$^{b}$, J. L\'opez-Sarri\'on$^{b}$,
  F. M\'endez$^{b}$ and P. A. G. Pisani$^a$}
  \email{falomir@fisica.unlp.edu.ar, jgamboa55@gmail.com, justinux75@gmail.com, feritox@gmail.com,
  pisani@obelix.fisica.unlp.edu.ar}
\affiliation{$^{a}$IFLP/CONICET   -   Departamento   de   F\'{\i}sica,
  Facultad de Ciencias Exactas,
Universidad Nacional de La Plata, C.C. 67, (1900) La Plata, Argentina}
\affiliation{$^{d}$Departamento   de   F\'{\i}sica,   Universidad   de
  Santiago de Chile, Casilla 307, Santiago, Chile}

\begin{abstract}
A general three-dimensional noncommutative quantum mechanical system mixing spatial and spin degrees of freedom is
proposed. The analogous of the harmonic oscillator in this description  contains a magnetic dipole interaction and  the ground state is explicitly computed and we show that
it is infinitely degenerated and implying a spontaneous symmetry breaking. The model can be straightforwardly extended to many particles and the main above properties are retained. Possible applications to the Bose-Einstein condensation with dipole-dipole interactions are briefly discussed.

\end{abstract}
\maketitle

In the last few years several aspects of noncommutative quantum mechanics (NCQM) have been
discussed extensively from different points of view \cite{NCQM}.  One of the main motivations was to consider NCQM as a
theoretical laboratory where some ideas of quantum field theory could be realized and --at the same time- to develop NCQM
as a new calculational tool in standard quantum mechanics.

In this paper we would like to propose a non-relativistic NCQM model where dipole (and higher multipole) interactions are
generated as a consequence of a particular commutative realization of the noncommutative algebra. In so doing, we
will first consider a general system where the above mentioned realization is presented and after that the analogous of the
harmonic oscillator will be explicitly discussed.

This last example is  non-trivial  since in the
experiments  the atoms are confined to a finite region of the space,
situation which is usually modellated, in a first approximation,  by
the harmonic oscillator. In addition, the present model for potentials
depending on higher powers of coordinates, introduces also spin-spin
interactions. This fact is relevant, for example, for 
Bose-Einstein condensation experiments with  ultracold
$^{52}$Cr \cite{german, german1}. The bosonic chromium isotopes have a vanishing nuclear spin, very large magnetic moment and integer total spin.  In
this context the dipole-dipole interactions become dominant if one adjust suitably the coefficients in the contact
interaction and, therefore --in this regime--  one could study  new
and interesting properties by using our model.

Thus, although many approaches are consecrated to dipolar systems \cite{review}, one the goals of this paper will be offer  a new and unconventional approach to systems.

In order to expose our results, let us consider the non-commutative spatial coordinates of a three-dimensional space,
${\hat x}_i$, the conjugate momenta, $\hat{p}_i$, and spin variables, ${\hat s}_i$, which we will assume to satisfy the non-
standard deformed Heisenberg algebra
\eqb
\left[{\hat x}_i,{\hat x}_j\right] &=& i\theta^2 \epsilon_{ijk} {\hat s}_k, \nonumber
\\
\left[{\hat x}_i,{\hat p}_j\right] &=& i \delta_{ij}, \,\,\,\,\,\,\,\,\,\,\,\,\,\,\,\,\,\,\,\,\,\,\,\left[{\hat p}_i,{\hat p}_j\right] = 0, \label{3}
\\
\left[{\hat x}_i,{\hat s}_j\right] &=& i \theta \epsilon_{ijk} {\hat s}_k,  \,\,\,\,\,\,\,\,\,\,\,\,
\left[{\hat s}_i,{\hat s}_j\right] = i \epsilon_{ijk} {\hat s}_k, \nonumber
\eqf
where $\theta$ is a parameter with length dimension and the indices $i,j,k$ run from $1$ to $3$.

The first commutator is reminiscent of the Snyder's algebra \cite{snyder}, who considered for the first time  a noncommutative Lorentz-invariant spacetime (see also \cite{
cnyang}). In our non-relativistic model, instead of closing the coordinates algebra to the total angular momentum components
--as required by Lorentz-invariance in \cite{snyder}-- we will consider only the spin operator (${\bf s}$). Otherwise, one should
also modify the commutator $[x,p]$ in order to satisfy the Jacobi's identity \cite{italia}.

The algebra (\ref{3}) can be explicitly realized in terms of \emph{commutative} variables by means of the identification
\eqb
{\hat x}_i  && \rightarrow {\hat x}_i = x_i + \theta s_i, \nonumber
\\
{\hat p}_i && \rightarrow {\hat p}_i = p_i:= - \imath \partial_i, \label{6}
\\
{\hat s}_i && \rightarrow {\hat s}_i = s_i:=\frac{\sigma_i}{2}, \nonumber
\eqf
where $x_i$ and $p_i$ are now canonical operators satisfying the Heisenberg's algebra. Notice the matricial character of the
non-commutative coordinate operators ${\hat x}_i$.

This simple observation implies that any noncommutative quantum mechanical system described by
\bb
\imath \partial_t\left| \psi(t) \right\rangle =
{\hat H} ({\hat p},{\hat x},{\hat s})  \left|\psi(t)\right\rangle 
= \left[ \frac{1}{2} {\hat p}^2 + {\hat V} ({\hat x})
\right] \left|\psi(t) \right\rangle
\label{8}
\ee
can  equivalently be described by the \emph{commutative} Schr\"odinger equation
\bb
\imath \partial_t \psi (\mathbf{x},t)= H(p_i, x_i + \theta s_i)\, \psi (\mathbf{x},t) \,,
\label{9}
\ee
where $\psi (\mathbf{x},t)$ is a Pauli spinor.

The quantum mechanical system (\ref{9}) cannot be  solved exactly for a general Hamiltonian, but one can consider particular
examples from which one can try to extract more general conclusions. Indeed, let us consider as an illustrative example the
isotropic harmonic oscillator in this three-dimensional non-commutative space, described by the Hamiltonian
\eqb
H &=& -\frac{1}{2} \nablavec^2 + \frac{1}{2} {\hat  \x}^2, \nonumber
\\
&=&-\frac{1}{2} \nablavec^2+ \frac{1}{2} \left({\x} + \theta {\bf s} \right)^2. \label{99}
\eqf
Contrarily to the commutative case, this is a non-trivial example due the presence of the ${\bf x}.{\bf s}$  \emph{dipole}
interaction, whose effects could be incorporated through perturbation theory.

Instead of studying this system, we will consider the supersymmetric version of this model, whose ground state can be
found  by means of the usual supersymmetry techniques. Indeed, notice that the Hamiltonian (\ref{99}) can be
written as
\bb
{\tilde H}= H-E_0= Q^\dagger_i Q, \label{10}
\ee
where ${\tilde H}$ is the commutative Hamiltonian with the ground state energy of the (commutaive) isotropic harmonic
oscillator
subtracted ($E_0=3/2$), and
\bb
Q_i = \frac{1}{\sqrt{2}} \left( \partial_i + {\hat x}_i\right), ~~~~~~~~~~Q^\dagger_i = \frac{1}{\sqrt{2}} \left(
-\partial_i +
{\hat x}_i\right), \label{11}
\ee
where ${\hat x}_i = x_i +\theta s_i$.

In the conventional (commutative) case, $Q_i$ and $Q^\dagger_i$ are just creation and annihilation operators, but
in the present non-commutative case this interpretation is lost and the explicit calculation of the commutators gives
\eqb
\left[Q_i,Q_j\right] &=& \frac{i \theta^2}{2}\epsilon_{ijk}s_k = \left[Q^\dagger_i,Q^\dagger_j\right],
\nonumber
\\
\left[Q_i,Q^\dagger_j\right] &=& \delta_{ij} + \frac{i \theta^2}{2} \epsilon_{ijk}s_k \label{12}
\eqf
(which reduces to the standard algebra of creation and annihilation operators in the $\theta \rightarrow 0$ limit).

\medskip

Following references \cite{gz} and \cite{dfgm} (see also \cite{gozzi}), we construct a supersymmetric version of the above
system by defining supercharges following the Ansatz,
\eqb
Q_i \rightarrow S&=& Q_i\psi_i=Q_i \sigma_i \otimes \sigma_-
=Q\otimes \sigma_-, \label{re1}
\\
Q^\dagger_i \rightarrow S^\dagger &=& Q^\dagger_i\psi^\dagger_i=Q^\dagger_i \sigma_i \otimes \sigma_+
=Q^\dagger \otimes \sigma_+, \label{
re2}
\eqf
where $Q=Q_i \sigma_i$, $Q^\dagger=Q^\dagger_i \sigma_i$ and  $\sigma_\pm = \frac{1}{2}(\sigma_1 \pm i \sigma_2)$, with
the property $\sigma^2_\pm =0$.

The supersymmetric Hamiltonian is defined as the non-negative operator
\begin{equation}\label{Ham-SUSY}
    H_s:= \frac{1}{2}\{S^\dagger, S\}=
    \frac{1}{2} \,Q^\dagger Q \otimes \frac{\mathbf{1+\sigma_3}}{2}
    +\frac{1}{2}\, Q Q^\dagger \otimes \frac{\mathbf{1-\sigma_3}}{2}\,,
\end{equation}
and this prescription obviously fulfills
\eqb
\left[ S,H_s\right] &=&0=\left[ S^\dagger,H_s\right], \label{ha2}
\\
\{S,S\}&=& 0 = \{S^\dagger, S^\dagger \}\,. \nonumber
\eqf
Notice that this Hamiltonian is defined on a space of four component functions (a pair of spinors),
$$
\psi =
\left( \begin{array}{c} \Psi^{(1)} \\
\Psi^{(2)}\end{array}\right)\,.
$$
A straightforward calculation yields
\eqb
H_s &=& \frac{1}{2}\, \left( -\frac{1}{2} \nabla^2 +\frac {1}{2} {\bf x}^2 +3~\theta ~{\bf x} \cdot {\bf s} + \frac{9}{8} \theta^2 \right)
\otimes {\openone}_{2\times 2} -\frac{1}{2}\,
\left(2\, {\bf s} \cdot {\bf L}+\frac{3}{2}\right) \otimes \sigma_3, \nonumber
\\
&=& H^{ss} +H_{NC}, \label{su}
\eqf
where
\bb
H^{ss}=\frac{1}{2}\, \left( -\frac{1}{2} \nabla^2 +\frac {1}{2} {\bf x}^2 \right) \otimes {\openone}_{2\times 2} -  \frac{1}{2}\,
\left(2\, {\bf s} \cdot {\bf L}+\frac{3}{2}\right) \otimes
\sigma_3, \label{osc1}
\ee
is the standard supersymmetric Hamiltonian in three-dimensions for the harmonic oscillator, whereas
\bb
H_{NC} = \frac{1}{2}\,\left(3~\theta ~{\bf x} \cdot {\bf s}  + \frac{9}{8} \theta^2\right) \otimes {\openone}_{2\times 2} , \label{nc1}
\ee
is the correction due to non-commutativity. Actually, the term ${\bf x}.{\bf s}$ is the dipole interaction mentioned above and
$\frac{9}{8} \theta^2$ is just a correction to the ground state energy.

One can extract physical information about this noncommutative supersymmetric oscillator by noticing that, from (\ref{Ham-SUSY}), it follows that the ground state satisfy
$$
S \psi_0 =0 \qquad {\rm or} \qquad S^\dagger \psi_0=0\,.
$$
For example, if $S \psi_0 =0$ then
$$\left[Q_i \sigma_i \otimes \sigma_- \right] \psi_0=0,$$
which implies that
$\psi_0=\left(
                           \begin{array}{c}
                             \Psi_0^{(1)} \\
                             \mathbf{0} \\
                           \end{array}
                         \right)
$, with $\Psi_0^{(1)}$ a spinor satisfying
\bb
Q\, \Psi_0^{(1)} = Q_i \sigma_i \Psi_0^{(1)} =0\,.  \label{i1}
\ee
It can be seen that the general normalized solution of this equation is
\bb
\Psi_0^{(1)} = \pi^{-\frac{3}{4}}
\, e^{\frac{3\theta}{2}\,\imath\, {\bf k}_I\cdot {\bf x}}
\, e^{-\frac{1}{2} \left({\bf x} -\frac{3\theta}{2} \mathbf{k}_R\right)^2} \chi_- (\hat{{\bf k}}), \label{gst}
\ee
where ${\hat {\bf k}}= {\bf k}_R +i {\bf k}_I$ is a complex unitary vector (
$\hat{\mathbf{k}}^2=1\ \Rightarrow \ {\mathbf{k}_R}^2-{\mathbf{k}_I}^2=1\,, \ \mathbf{k}_R \cdot \mathbf{k}_I=0$) and
$\chi_- (\hat{\mathbf{k}})$ is a constant spinor satisfying
\begin{equation}
    \left({\hat {\bf k}} \cdot \sigmavec \right)  \chi_-({\hat {\bf k}}) = -\chi_- ({\hat {\bf k}}), \nonumber
\end{equation}
with $\chi_-(\hat{\mathbf{k}})^\dagger \chi_-(\hat{\mathbf{k}})=1$. On the other hand, it is easy to see that
$S^\dagger \psi_0=0$ has no normalizable solutions.

From  (\ref{gst}) we see that the ground state is infinitely degenerated and that the rotational symmetry is spontaneously
broken.
Indeed, it is straightforward to get for the mean value of $\mathbf{x}$ and $\mathbf{p}=-\imath \nablavec$ in this ground state
\begin{equation}\label{x-medio}
    \langle \mathbf{x} \rangle_{\mathbf{k}}=
    \left(\Psi^0_{\mathbf{k}} ,\mathbf{x}\, \Psi^0_{\mathbf{k}} \right)=
    \pi^{-\frac{3}{2}}\int d^3x\, e^{-\left( \mathbf{x}-\frac{3\theta}{2}\mathbf{k}_R \right)^2} \mathbf{x}
    = \frac{3\theta}{2}\,\mathbf{k}_R\,,
\end{equation}
\begin{equation}\label{p-medio}
    \langle \mathbf{p} \rangle_{\mathbf{k}}=
    \left(\Psi^0_{\mathbf{k}} ,-\imath \nablavec\, \Psi^0_{\mathbf{k}} \right)=
    {-\imath\, \pi^{-\frac{3}{2}}}\int d^3x\, e^{-\left( \mathbf{x}-\frac{3\theta}{2}\mathbf{k}_R \right)^2} \left[\frac{3\theta}{2}\, \imath \, \mathbf{k}_I-
    \left(\mathbf{x}-\frac{3\theta}{2}\mathbf{k}_R\right)\right]= \frac{3\theta}{2}\,\mathbf{k}_I\,.
\end{equation}
Therefore, the real part of $\hat{\mathbf{k}}$ determines the departure of the mean position of the particle from the origin,
while the imaginary part determines the mean linear momentum of the particle in the ground state. Notice that the departure
from the origin grows with the mean linear momentum in a direction orthogonal to it.

Similarly, the mean value of the orbital angular momentum in the ground state is easily obtained as
\begin{equation}\label{L-medio}
    \begin{array}{c}
    \displaystyle
      \left(\Psi^0_{\mathbf{k}} ,\mathbf{L}\, \Psi^0_{\mathbf{k}} \right)=
    {-\imath\, \pi^{-\frac{3}{2}}}\int d^3x\,
    e^{-\frac{3\theta}{2}\,\imath\, {\bf k}_I\cdot {\bf x}}
    \, e^{-\frac{1}{2} \left({\bf x} -\frac{3\theta}{2} \mathbf{k}_R\right)^2}
    \left(\mathbf{x} \times \nablavec\right) e^{\frac{3\theta}{2}\,\imath\, {\bf k}_I\cdot {\bf x}}
    \, e^{-\frac{1}{2} \left({\bf x} -\frac{3\theta}{2} \mathbf{k}_R\right)^2}=
    \\ \\ \displaystyle
      =-\imath \, \frac{3 \theta}{2 \pi^{3/2}}
      \int d^3x\, e^{-\left( \mathbf{x}-\frac{3\theta}{2}\mathbf{k}_R \right)^2}
      \left(\mathbf{x} \times \mathbf{k}\right) =
      \left(\frac{3 \theta}{2}\right)^2 \, \mathbf{k}_R \times \mathbf{k}_I
      = \langle \mathbf{x} \rangle_{\mathbf{k}} \times \langle \mathbf{p} \rangle_{\mathbf{k}}\,,
    \end{array}
\end{equation}
which also depends continuously on the vector $\mathbf{k}$.

On the other hand, since the energy of the ground state vanishes, the supersymmetry of the model is manifest, and the
partner
Hamiltonians, $H_+=\frac{1}{2}\, Q^\dagger Q$ and $H_-=\frac{1}{2}\,Q\,Q^\dagger$, share the same spectrum except for the
ground state.

\medskip

These characteristics of our model are quite nontrivial properties, and are reminiscent of the spontaneous symmetry
breaking phenomenon in quantum field theory. Actually,  as is well known, the spontaneous symmetry breaking of the
vacuum for a scalar field theory is characterized by a vacuum expectation value $<0|\phi |0>= \mbox{const.}\neq 0$. Thus,
such as in quantum field theory, one sees that the spontaneous symmetry breaking phenomenon is also present in our
example.

Following the above analogy with quantum field theory one could argue that the presence of dipolar interactions leads
to phase transitions in the same sense as the spontaneous symmetry breaking vacuum yields to phase transitions, {\it e.g.}
in mean field theory (Ginzburg-Landau).

The extension of the above model to many interacting particles is straightforward, {\it e.g.}
\bb
H = \sum_{i=1}^N \frac{{\bf p}_i^2}{2m} + \frac{\lambda_1}{2} \sum_{i\neq j} \left({\hat x}_i -{\hat x}_j\right)^2 +
\frac{\lambda_2}{2} \sum_{i\neq j} \left({\hat x}_i -{\hat x}_j\right)^4+ ... \label{hami1}
\ee
where $i,j$ denotes particles indices, and the appearance of dipolar, quadrupolar and higher multipole interactions are direct
to see.

In conclusion, in this letter we have proposed an approach to the magnetic multipolar interactions based on a model of NCQM that contain very peculiar properties such as spontaneous symmetry breaking and infinite degenerate ground states. We suggest that these properties continue being valid even though new interactions be included. Indeed, as was shown above, quadratic interactions induce a nontrivial vacuum structure and, therefore, for $N$ particles a good ground state could be the trial function
\bb
\psi ({\bf x}_1, {\bf x}_2, ...) = \psi ({\bf x}_1).\psi ({\bf x}_2).\psi ({\bf x}_3)...
\ee
where each $\psi ({\bf x})$ correspond to the harmonic oscillator ground state discussed above.

Within this model, a system dominated by dipole interactions --as the $^{52}$Cr gas-- would have infinitely degenerated ground states, with a spontaneous symmetry breakdown associated to a transition isotropic-anisotropic, {\it i.e.} possibly related to a new phase.

\vspace{0.3 cm}

\noindent\underline{Acknowledgements}:  We would like to thank to J. C. Retamal and M. Loewe by fruitful discussions.
This work was partially supported by FONDECYT-Chile grant-1095106, 1060079 and DICYT (USACH), and by CONICET (PIP
01787) and UNLP (Proy.\ 11/X492), Argentina.

\end{document}